\documentclass[%
 reprint,
superscriptaddress,
 amsmath,amssymb,
prapplied,
]{revtex4-2}

\usepackage{graphicx}% Include figure files
\usepackage{dcolumn}% Align table columns on decimal point
\usepackage{bm}% bold math
\begin{document}

\preprint{APS/123-QED}

\title{Deterministic loading of a single strontium ion into a surface electrode trap \\ using pulsed laser ablation}% Force line breaks with \\

\author{Alto Osada}
\email{alto@g.ecc.u-tokyo.ac.jp}
\affiliation{Komaba Institute for Science (KIS), The University of Tokyo, Meguro-ku, Tokyo, 153-8902, Japan }%Lines break automatically or can be forced with \\
 \affiliation{PRESTO, Japan Science and Technology Agency, Kawaguchi-shi, Saitama, 332-0012, Japan}%

\author{Atsushi Noguchi}
\affiliation{Komaba Institute for Science (KIS), The University of Tokyo, Meguro-ku, Tokyo, 153-8902, Japan }%Lines break automatically or can be forced with \\
\affiliation{
RIKEN Center for Quantum Computing (RQC), RIKEN, Wako-shi, Saitama, 351-0198, Japan
}
\affiliation{Inamori Research Institute for Science (InaRIS), Kyoto-shi, Kyoto 600-8411, Japan}

\date{\today}% It is always \today, today,

\begin{abstract}
Trapped-ion quantum technologies have been developed for decades toward applications such as precision measurement, quantum communication and quantum computation. 
Coherent manipulation of ions' oscillatory motions in an ion trap is important for quantum information processing by ions, however, unwanted decoherence caused by fluctuating electric-field environment often hinders stable and high-fidelity operations..
One way to avoid this is to adopt pulsed laser ablation for ion loading, a loading method with significantly reduced pollution and heat production.
Despite the usefulness of the ablation loading such as the compatibility with cryogenic environment, randomness of the number of loaded ions is still problematic in realistic applications where definite number of ions are preferably loaded with high probability.
%The ablation loading is proven to be useful, being even compatible with cryogenic environment, except for the randomness of the number of loaded ions.
In this paper, we demonstrate an efficient loading of a single strontium ion into a surface electrode trap generated by laser ablation and successive photoionization. 
The probability of single-ion loading into a surface electrode trap is measured to be 82\,\%, and such a deterministic single-ion loading allows for loading ions into the trap one-by-one. 
Our results open up a way to develop more functional ion-trap quantum devices by the clean, stable, and deterministic ion loading.
\end{abstract}

\maketitle

%\tableofcontents

\section{Introduction}

Trapped ion systems, in which atomic ions are levitated in a vacuum by electric and/or magnetic field, are promising platform for the development of quantum technologies such as quantum metrology~\cite{Dehmelt1982-hh,Burt2021-no,Brewer2019-jh}, quantum communication~\cite{Connell2017-yg,Matthias_Keller_Birgit_Lange_Kazuhiro_Hayasaka_Wolfgang_Lange_Herbert_Walther2004-ib,Schupp2021-os} and quantum computation~\cite{Cirac1995-bz,Brown2016-me}.  
Precise quantum operations on trapped ions as quantum bits (qubits) are indispensable ingredients for such a purpose.  
In particular, motions of ions work not only as a mediator of two-qubit gate~\cite{Molmer1999-dq,Gaebler2016-rd,Ospelkaus2011-go} but also a platform for encoding a qubit~\cite{Gottesman2001-rn,Fluhmann2019-ki}, and hence it is getting more and more important to precisely control them.

To date, it is known that surface contamination of trapping electrodes may result in unwanted heating of ions' motion~\cite{Brownnutt2015-oq} and/or in formation of patch potential that significantly changes the ions' position.  
In this regard, conventional methods of generating atomic ions such as electron impact ionization and photoionization combined with an atomic oven~\cite{Schioppo2012-cf} result in significant degradation of ultra-high vacuum environment or difficulty in heat management. 
A cleaner method of ion loading technique with minimal thermal footprint is anticipated both in terms of preserving ions' motional coherence and long-term stability of the system.

Pulsed laser ablation provides a cleaner way of generating neutral atoms and even atomic ions directly~\cite{Olmschenk2017-mf,Piotrowski2020-kz,Zimmermann2012-gv,Sheridan2011-ck,Shao2018-ak,Sameed2020-ac,Leibrandt2007-au,Vrijsen2019-xr,Wu2021-nq}. 
It usually utilizes a focused nanosecond or shorter laser pulse to heat a target material locally and remove tiny amount of constituent elements, which are thrown out into surrounding space.
Either by capturing ejected atomic ions directly or by succeeding photoionization of neutral atoms and trapping atomic ions, pulsed laser ablation has been established as a promising technique for the ion loading, for the following reasons.
Since the thermal influence is nanosecond-short, it allows for avoiding the unstable thermal environment during the ion loading sequence accompanied to the use of the atomic oven.
Pulsed laser ablation is also advantageous in terms of the cleanness, since the atoms are only ejected for a very short duration to result in a significant reduction of contamination in the vacuum chamber and thus on the trap electrodes.
Another important merit is the possibility of number-controlled ion loading in which the number of loaded ions per ablation pulse can be controlled by varying pulse fluence and trapping potential. 
Deterministic loading of only a single ion by laser ablation is expected, which is thought to be useful in e.g. the quantum CCD architecture~\cite{Kielpinski2002-dy}, for its small footprint in the vacuum chamber and simplicity of implementation compared to those utilizing another laser-cooled neutral atomic cloud~\cite{Hill2004-vv,Schnitzler2009-qa,Schnitzler2010-op,Hanssen2006-te}.

So far, pulsed laser ablation is implemented in loading ions such as $^{40}$Ca$^+$~\cite{Shao2018-ak}, $^9$Be$^+$~\cite{Sameed2020-ac}, $^{88}$Sr$^+$~\cite{Leibrandt2007-au}, $^{174}$Yb$^+$~\cite{Vrijsen2019-xr} and other ions~\cite{Piotrowski2020-kz,Olmschenk2017-mf} mostly using a Q-switched YAG laser or its high harmonics. 
The probability of single-ion loading is concerned in Refs.~\cite{Leibrandt2007-au,Vrijsen2019-xr}, where it amounts only less than 20\,\%~\cite{Leibrandt2007-au} and about 50\,\%~\cite{Vrijsen2019-xr} where ablated atomic ions are directly loaded into the trap in the former and the latter suffers from the events that two or more ions are simultaneously loaded.  It is highly desirable to improve the probability of the single-ion loading by the pulsed laser ablation in the scope of the number-controlled, efficient loading in harmony with the trapped-ion quantum technologies. In practice, it can be installed in a cryogenic environment~\cite{Niemann2019-yk,Dubielzig2021-qh} and in a surface trap~\cite{Leibrandt2007-au,Vrijsen2019-xr} and hence a quantum CCD architecture.

In this article, we demonstrate pulsed-laser-ablation loading of $^{88}$Sr$^+$ into a surface electrode trap with high single-ion-loading efficiency.
With SrTiO$_3$ as a long-lived ablation target~\cite{Leibrandt2007-au}, a Q-switched 1064~nm laser is used for it and proven to be valid for the laser ablation with its pulse energy being only several tens of $\mathrm{\mu}$J, nonetheless the target material is transparent in this wavelength region.
By varying the laser fluence, we observed that the number of loaded ions per ablation pulse changes as well.
In particular, we can load a single ion with high probability up to 82\,\%.
On top of these, we successfully load ions one-by-one into the surface electrode trap.
Our results manifest the ablation loading of ions as a possible new standard and opens up a way to various applications in quantum technologies with cleaner, heat-management-free, on-demand and deterministic ion preparation.

\begin{figure}[t]
\includegraphics[width=8.4cm]{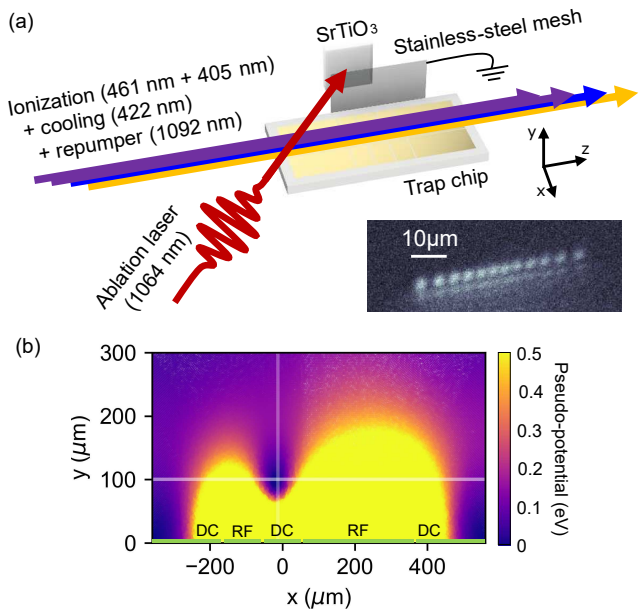}
\caption{\label{Fig1} (a) Schematic illustration of our experimental setup. A SrTO$_3$ plate is placed next to the trap chip and a stainless-steel mesh is inserted in between them. Lasers for the photoionization and laser cooling pass through slightly above the trap chip and points at the trapping region. Ablation laser is incident on the SrTiO$_3$ with an oblique angle of about 34 degree. Inset in the right-bottom displays an fluorescence image of twelve strontium ions levitated above the trap chip. (b) Pseudo-potential landscape of our surface trap calculated by finite element method with RF amplitude of 90~V.  Green bars at the bottom of the plot depict the electrodes, where $x=0$ corresponds to the middle of the central DC electrode. The crossing point of the horizontal and vertical white lines indicates the trapping center. Color scale in the figure is intended to saturate at 0.5~eV so that the trapping potential can be seen.}
\end{figure}

This article is constructed as follows. In Sec.~\ref{setup}, we introduce our experimental setup including a surface-electrode ion trap and pulsed laser ablation.
Section~\ref{results} describes experimental results including generation of the neutral strontium atoms, single-ion loading and one-by-one ion loading. Sec.~\ref{conclusion} concludes this article.

\section{Experimental setup} \label{setup}

\subsection{Experimental configurations}

A schematics of the experimental configuration is depicted in Fig.~\ref{Fig1}(a). A surface electrode trap chip, being made of electroplated alumina substrate with patterned gold electrodes, and a SrTiO$_3$ plate as an ablation target are placed inside the ultrahigh-vacuum ($\sim 4 \times 10^{-8}$~Pa) shroud.
SrTiO$_3$ is a good ablation target for generating strontium atoms since it keeps ablated and generating sufficient amount of strontium atoms even when it is laser-ablated tens of thousands of times by a pulsed 355~nm-wavelength laser~\cite{Leibrandt2007-au}. 
Moreover, the metallic strontium, which is frequently used in the oven method, is highly reactive to atmosphere, so that the handling of it and loading it into the vacuum shroud requires special care. 
Laser ablation of SrTiO$_3$ also solves this problem and makes strontium generation far more facile.
An electrically grounded stainless-steel mesh is put in between the two to block unwanted charged particles generated in the pulsed laser ablation.  
For the isotope-selective photoionization of the generated neutral strontium atoms, two lasers of wavelengths 461~nm ($^1S_{0} \,\leftrightarrow\, ^1P_{1}$) and 405~nm ($^1P_{1} \,\leftrightarrow\, ^1D_{2}$, $^1D_{2}$ is auto-ionizing) are needed and laser cooling of $^{88}$Sr$^+$ requires two additional lasers of 422~nm and 1092~nm wavelength, respectively for driving cooling ($^2S_{1/2}\, \leftrightarrow\, {^2}P_{1/2}$) and repumping ($^2D_{3/2}\, \leftrightarrow\, {^2}P_{1/2}$) transitions.
%All four laser beams are loosely focus down to 300~$\mu$m-waist with optical power being 7~mW for 405~nm, 4~mW for 461~nm, 1~mW for 422~nm and 3~mW for 1092~nm. 

All the four lasers are overlapped together using beamsplitters and dichroic mirrors.
They are aligned approximately parallel to $z$-axis as indicated in the figure, where axial motion of the ions in the surface electrode trap is aligned in $z$-axis as well.  
Pulsed 1064~nm laser impinges on the surface of SrTiO$_3$ plate with oblique incidence without touching the stainless-steel mesh. Details of the ablation laser will be described in Sec.~\ref{ablationlaser}.

As the laser pulse ablates SrTiO$_3$, only neutral particles including strontium atoms are intended to go through the mesh and flies into the potential minimum of the surface electrode trap.
Then only $^{88}$Sr atoms are selectively photoionized to become singly-ionized $^{88}$Sr$^+$, which starts to be laser-cooled in the trap. 
For reference, Fig.~\ref{Fig1}(b) displays the pseudo-potential landscape in $xy$-plane at the trapping region, calculated by finite-element method. 
Our trap adopts a standard five-rail electrode configuration~\cite{Hong2016-mp}, where one of the RF electrode is three times wider in $x$ direction. 
As a result, The trap potential is tilted in $xy$-plane by 15 degree and the trap center is shifted from the middle of the central DC electrode, as indicated by the crossing point of two white lines in Fig.~\ref{Fig1}(b).
Combined with slightly tilted laser beams in $yz$-plane, not only the motion in the $z$-axis is laser-cooled efficiently but those in $x$ and $y$ axes can be, though inefficiently, cooled down as well.
With applied 30.7~MHz RF of typically 90~V amplitude, depth of the trap reads 150~meV and secular frequencies become around 2.8~MHz. 
We can trap ions even with 45~V-amplitude RF that yields the trap depth of 39~meV. 
Varying the trap depth between these values does not have significant effect on the single-ion loading experiment, the main scope of this work.
Typical optical microscope image of the fluorescence from laser-cooled trapped ions are shown in the right-bottom of Fig.~\ref{Fig1}(a) where a one-dimensional Wigner crystal containing twelve 88-strontium ions are visible.

\subsection{Pulsed laser for ablation} \label{ablationlaser}

As an ablation laser, Q-switched 1064~nm-wavelength laser is used. 
The maximum available single-pulse energy is $300$~$\mu$J and the pulse width is 6~ns which are verified with separate measurements. 
In the experiment, we set the pulse energy to be around 100~$\mu$J that is revealed to be enough for the laser ablation of SrTiO$_3$. 
Our pulsed laser is originally prepared for the laser ablation of metallic strontium and it actually works, that is, the strontium atoms are generated.
However, through tens of ablation pulses the yield of strontium atoms rapidly degrades.
Moreover, handling problem hinders the clean installation and reproducible investigations.  
We tried laser ablation of SrTiO$_3$ with this near-infrared pulsed laser and found that this combination of the laser and the material works well, although being surprising because SrTiO$_3$ is highly transparent in near-infrared wavelength.
Though the reason for this success is unclear, multi-photon processes due to the relatively large fluence and short pulse duration may enable the absorption of 1064~nm light by the SrTiO$_3$ and hence the local, short-time heating~\cite{Jones1989-hw}. 

As shown in Fig.~\ref{Fig1}(a), the ablation laser is shined on the target SrTiO$_3$ plate next to the trap chip at an oblique angle.
Owing to this configuration, the ablation laser is not disturbed by the stainless-steel mesh while the ablated particles of various types directed toward the trapping region go through the mesh, by which a fraction of unwanted charged particles is filtered out.
The angle between the optical axis of the ablation laser and $x$-axis is about 34 degree, resulting in a slightly oblate elliptical beam spot at SrTiO$_3$ surface with minor diameter of 50~$\mu$m and major one being 60~$\mu$m focused by a spherical lens with the focal length of 250~mm. Therefore, the single-pulse laser fluence typically adopted in our experiment is around 5~J/cm$^2$. 
Throughout the experiment, we observed no degradation of the ultrahigh vacuum $\sim 4\times 10^{-8}$~Pa with the ablation laser fluence of less than 6~J/cm$^2$, confirming that the ablation loading of ions is far cleaner than other methods adopting ovens or high-flux electron beams. 
%However, with 9~J/cm$^2$ fluence, the vacuum degrades to $\sim 3\times 10^{-7}$~Pa for several minutes and then gradually gets back to the original value.

\begin{figure}[t]
\includegraphics[width=8.4cm]{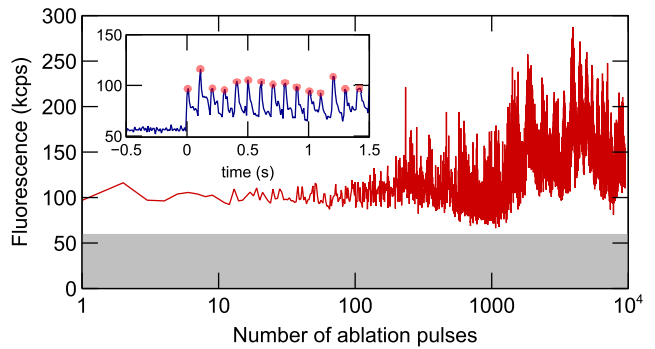}
\caption{\label{Fig2} Observed 422~nm-wavelength fluorescence counts taken at laser ablation after a various number of repeated laser ablation pulses (red plot). The upper bound of the gray shaded area indicates the count level when the ablation laser pulse is turned off and the trap is empty.  Inset: Fluorescence counts recorded as a function of time during the ablation laser pulses are applied at 10~Hz repetition rate. The ablation laser pulse train starts at 0~s. Red dots represent first fifteen peak values used in the red plot.}
\end{figure}

\section{Results} \label{results}

\subsection{Generation of neutral strontium atoms}

For the generation and observation of neutral strontium atoms, the laser frequencies are tuned properly, the laser beams are aligned so that they point at the trap center, and RF and DC voltages are applied to the surface electrode trap.  Then we fire the ablation pulse on the target and investigate the behavior of ions loaded into the trap. 
Firstly we examine the amount of generated strontium atoms by collecting the 422~nm fluorescence from the trapping region detected by a photomultiplier tube.
Fluorescence photons are collected through an objective lens with its numerical aperture of 0.3.
We make a brief note on why we record the fluorescence of strontium ions to estimate the amount of generated neutral strontium atoms. 
The neutral strontium atoms ablated out of the ablation target goes through the mesh in front of the target and reaches the trap region. They are photoionized by the combination of 461~nm- and 405~nm-wavelength lasers and laser cooling of ion cloud follows.
These processes occur rapidly in the trapping region and basically the amounts of temporarily trapped ions and generated neutral strontium atoms can be thought of as being proportional to each other, and so the detected photocounts at the photomultiplier tube and the number of generated neutral strontium atoms~\cite{Leibrandt2007-au}.

The inset of Fig.~\ref{Fig2} shows time evolution of the detected fluorescence at the photomultiplier tube with ablation laser pulses starting at 0~s with repetition rate of 10~Hz. The single-pulse fluence is 9.0~J/cm$^2$ here which is relatively intense and hence the white light-emitting spot blinking at the ablated region is visible to naked eye.  This originates in the recombination of the electrons and ions contained in the ablation plume.
In the blue plot, peaks are visible at every time an ablation laser pulse impinges on the ablation target.
This signal structure is saw-tooth shaped since, with the parameters in this experiment, the laser cooling is intentionally made not so effective that loaded ions can escape the trap to result in the quickly decreasing fluorescence counts.
In this manner, we record the peak fluorescence count (indicated by red dots) at each ablation event and present the result shown as a red plot in Fig.~\ref{Fig2}.
The upper bound of the gray shaded area represents the signal level when the ablation laser pulse is blocked.
The fluorescence counts always record values above this baseline and no tendency of degradation can be seen even after $10\,000$ ablation events.
This result is consistent with the one described in Ref.~\cite{Leibrandt2007-au}.
Our near-infrared pulsed laser ablation may have even better performance than that in  Ref.~\cite{Leibrandt2007-au} where an ultraviolet pulsed laser ablation is investigated.
However, more systematic study should take place, which is beyond the scope of this article.

\begin{figure}[t]
\includegraphics[width=8.4cm]{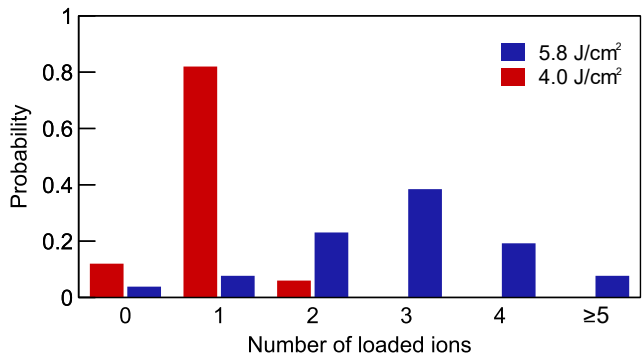}
\caption{\label{Fig3} Probabilities of loading 0 to 4 and more ions into the trap. Green, red and blue ones correspond to those measured with 4.0~J/cm$^2$ and 5.8~J/cm$^2$ fluence of the ablation laser pulse with 50 and 26 attempts, respectively. }
\end{figure}

\subsection{Efficient loading of a single $^{88}$Sr$^+$} 

As shown in the previous Section, our ablation scheme works well for the ion loading. 
Once the trapping parameters are tuned so that efficient laser cooling of ions is achieved, we investigate the distribution of the number of loaded ions.
Results are displayed as histograms in Fig.~\ref{Fig3}. The vertical axis represents the probability and red and blue data are experimentally measured probability distributions respectively for 4.0~J/cm$^2$ laser fluence with 50 attempts and for 5.8~J/cm$^2$ laser fluence with 26 attempts. 
Looking at the blue data with relatively strong 5.8~J/cm$^2$ laser fluence, the most frequent event was that three ions are loaded into the trap with measured probability of about 40\,\%. With comparable probabilities to this, two or four ions are loaded and fewer chances are given for loading less than two or more than four ions.
On the other hand, by decreasing the laser fluence down to 4.0~J/cm$^2$, the probability that only one ion is loaded amounts to 82\,\%, with greatly suppressed probabilities of other cases, see red plot in Fig.~\ref{Fig3}.
The single-ion-loading probability of $\sim$ 80\,\% is routinely obtained in the experiment with this laser fluence.
This distribution does not obey Poissonian statistics, in contrast to the experimental result in Ref.~\cite{Leibrandt2007-au} in which ablation loading of ions is done without the photoionization lasers. 
For further discussion about this issue, please refer to Sec.~\ref{discussion}.

\begin{figure}[t]
\includegraphics[width=8.4cm]{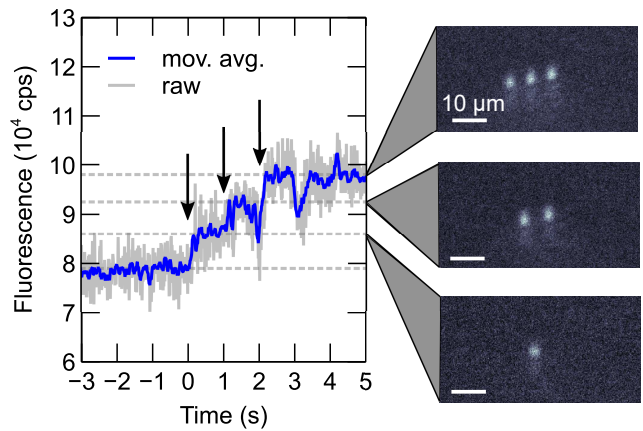}
\caption{\label{Fig4} Fluorescence counts recorded as a function time, where three ablation laser pulses are applied at 0~s, 1~s and 2~s (downward arrows in the figure). Gray plot represents the raw data and blue one is the moving average of it over 9 neighboring data points. Dotted horizontal lines indicate the signal levels of the situations that zero, one, two and three ions are present in the trap, as indicated with fluorescence images in the right panel.}
\end{figure}

With 4.0~J/cm$^2$ laser fluence with which the highly efficient single-ion loading is available, we demonstrate the loading of ions into the trap one by one.
Figure~\ref{Fig4} shows fluorescence counts as a function of time.
The ablation laser pulse is applied at 0~s, 1~s and 2~s in this experiment and the lasers for photoionization and laser cooling are always turned on.
The step-like increment of the fluorescence counts are visible around the times when the three ablation pulses impinges, and camera images containing one to three ions shown in the right panel verify that these confirms the loading of three ions one by one into the same trap.
There can be seen sudden decrements of fluorescence counts at about 1.8~s and 3~s. 
These are possibly due to the population trapping in $^{2}D_{3/2}$ state and/or meltdown of the ion crystal caused by fluctuating frequency of the repumper laser which is not frequency-stabilized to a reference in this experiment.
Though not being shown in the figure, the one-by-one loading of ions up to 6 ions are experimentally observed.
Since the repetition rate of the one-by-one loading can be as fast as sub-kHz, this loading scheme potentially enables us to prepare e.g. 5 ions with in 10~ms, which provides a fast and deterministic method of ion preparation.

\subsection{Discussion} \label{discussion}

There are some arguments on what kind of statistics do the distributions in Fig.~\ref{Fig3} obeys. 
For example, Leibrandt \textit{et al.}~\cite{Leibrandt2007-au} load ablated ions directly into the trap and observed the Poissonian distribution of the number of loaded ions.
Other experiments such as Ref.~\cite{Wu2021-nq, Vrijsen2019-xr} revealed that the distribution of the number of loaded ions were not Poissonian in their cases.
Wu \textit{et al.} claims that Levi distribution fitted their results well, though the reason is still unclear.
Full loading dynamics is difficult to simulate, where many interacting ions move rapidly in the RF electric field and the laser cooling requires $\sim$1~s to make equilibrium ion chain.
Our results in Fig.~\ref{Fig3} also suggest the non-Poissonian nature of the number distribution, suggesting that such a few-body or many-body dynamics in the RF electric field might have some effect.
To extract some factors that enables the efficient single-ion loading, we list the parameters for ablation loading experiments in Table~\ref{tb}.
The single-ion-loading probabilities higher than 0.5 are obtained so far in the loading scheme utilizing photoionization for the ion generation and the surface trap in which typical trap depth is about 100~meV or less. 
Therefore, conditions of the laser cooling and trapping may also have an effect on the efficiency of the single-ion loading, which requires further investigation.

\begin{table}[t]
    %   \centering
      \begin{tabular}{cccccc}
         & $E_p$ ($\mu$J) & $p$ & $\lambda$ (nm) & PI & Misc.\\ \hline\hline
        % $^9$Be$^+$~\cite{Niemann2019-yk} & 40-80 &  & 532 & yes & Penning \\
        $^9$Be$^+$~\cite{Wu2021-nq} & 20 & $\sim$0.2 & 532 & no & Penning \\
        % $^{40}$Ca$^+$~\cite{Shao2018-ak} & 50 &   & 532 & yes & Surface trap \\
        $^{88}$Sr$^+$~\cite{Leibrandt2007-au} & 1100 & $\sim$0.15 & 355 & no & surface trap\\
        $^{174}$Yb$^+$~\cite{Vrijsen2019-xr} & 250 & $\sim$0.5 & 1064 & yes & surface trap\\
        This work & 100 & 0.82 & 1064 & yes & surface trap\\\hline
      \end{tabular}
      \caption{List of parameters and conditions of ablation-based single-ion-loading experiments. $E_p$, $p$ and $\lambda$ denote pulse energy, single-ion-loading probability and wavelength of the Q-switched laser, respectively. Whether the photoionization (PI) is implemented or not is also shown. As a miscellaneous information, type of the ion trap is accompanied as well.}
      \label{tb}
\end{table}

\section{Conclusion} \label{conclusion}

We demonstrated the highly efficient, single-ion loading of $^{88}$Sr$^+$ into the surface electrode trap. 
The 1064~nm-wavelength, nanosecond pulsed laser could be used for the laser ablation of SrTiO$_3$ target to produce neutral strontium atoms over 10$^4$ times and potentially more, without noticeable influence on thermal and vacuum environments.  
With 4.0~J/cm$^2$ fluence of the ablation laser pulse, we achieved the single-ion-loading probability of 82\,\% which made a step forward to the realization of deterministic single-ion loading into the surface trap.
Furthermore, we demonstrated the one-by-one loading of ions into the trap by applying desired number of ablation laser pulses.
Since the repetition rate of the one-by-one loading can be much faster, applying a chunk of multiple single-ion loading pulses might work as a fast and deterministic way of fixed-number ion loading.  Regarding the compatibility of the ablation loading with quantum CCD architecture and cryogenic environment, our results add an important ingredient for the further development of trapped-ion quantum technologies.

This work was supported by JST PRESTO (Grant No. 19198708), JSPS KAKENHI (Grant No. 21K13861), Murata Science Foundation, The Mitsubishi Foundation. AO acknowledge Ippei Nakamura and Masato Shigefuji for useful discussions.
\bibliography{main}

\begin{thebibliography}{33}%
\makeatletter
\providecommand \@ifxundefined [1]{%
 \@ifx{#1\undefined}
}%
\providecommand \@ifnum [1]{%
 \ifnum #1\expandafter \@firstoftwo
 \else \expandafter \@secondoftwo
 \fi
}%
\providecommand \@ifx [1]{%
 \ifx #1\expandafter \@firstoftwo
 \else \expandafter \@secondoftwo
 \fi
}%
\providecommand \natexlab [1]{#1}%
\providecommand \enquote  [1]{``#1''}%
\providecommand \bibnamefont  [1]{#1}%
\providecommand \bibfnamefont [1]{#1}%
\providecommand \citenamefont [1]{#1}%
\providecommand \href@noop [0]{\@secondoftwo}%
\providecommand \href [0]{\begingroup \@sanitize@url \@href}%
\providecommand \@href[1]{\@@startlink{#1}\@@href}%
\providecommand \@@href[1]{\endgroup#1\@@endlink}%
\providecommand \@sanitize@url [0]{\catcode `\\12\catcode `\$12\catcode
  `\&12\catcode `\#12\catcode `\^12\catcode `\_12\catcode `\%12\relax}%
\providecommand \@@startlink[1]{}%
\providecommand \@@endlink[0]{}%
\providecommand \url  [0]{\begingroup\@sanitize@url \@url }%
\providecommand \@url [1]{\endgroup\@href {#1}{\urlprefix }}%
\providecommand \urlprefix  [0]{URL }%
\providecommand \Eprint [0]{\href }%
\providecommand \doibase [0]{https://doi.org/}%
\providecommand \selectlanguage [0]{\@gobble}%
\providecommand \bibinfo  [0]{\@secondoftwo}%
\providecommand \bibfield  [0]{\@secondoftwo}%
\providecommand \translation [1]{[#1]}%
\providecommand \BibitemOpen [0]{}%
\providecommand \bibitemStop [0]{}%
\providecommand \bibitemNoStop [0]{.\EOS\space}%
\providecommand \EOS [0]{\spacefactor3000\relax}%
\providecommand \BibitemShut  [1]{\csname bibitem#1\endcsname}%
\let\auto@bib@innerbib\@empty
%</preamble>
\bibitem [{\citenamefont {Dehmelt}(1982)}]{Dehmelt1982-hh}%
  \BibitemOpen
  \bibfield  {author} {\bibinfo {author} {\bibfnamefont {H.~G.}\ \bibnamefont
  {Dehmelt}},\ }\href@noop {} {\bibfield  {journal} {\bibinfo  {journal} {IEEE
  Trans. Instrum. Meas.}\ }\textbf {\bibinfo {volume} {IM-31}},\ \bibinfo
  {pages} {83} (\bibinfo {year} {1982})}\BibitemShut {NoStop}%
\bibitem [{\citenamefont {Burt}\ \emph {et~al.}(2021)\citenamefont {Burt},
  \citenamefont {Prestage}, \citenamefont {Tjoelker}, \citenamefont {Enzer},
  \citenamefont {Kuang}, \citenamefont {Murphy}, \citenamefont {Robison},
  \citenamefont {Seubert}, \citenamefont {Wang},\ and\ \citenamefont
  {Ely}}]{Burt2021-no}%
  \BibitemOpen
  \bibfield  {author} {\bibinfo {author} {\bibfnamefont {E.~A.}\ \bibnamefont
  {Burt}}, \bibinfo {author} {\bibfnamefont {J.~D.}\ \bibnamefont {Prestage}},
  \bibinfo {author} {\bibfnamefont {R.~L.}\ \bibnamefont {Tjoelker}}, \bibinfo
  {author} {\bibfnamefont {D.~G.}\ \bibnamefont {Enzer}}, \bibinfo {author}
  {\bibfnamefont {D.}~\bibnamefont {Kuang}}, \bibinfo {author} {\bibfnamefont
  {D.~W.}\ \bibnamefont {Murphy}}, \bibinfo {author} {\bibfnamefont {D.~E.}\
  \bibnamefont {Robison}}, \bibinfo {author} {\bibfnamefont {J.~M.}\
  \bibnamefont {Seubert}}, \bibinfo {author} {\bibfnamefont {R.~T.}\
  \bibnamefont {Wang}},\ and\ \bibinfo {author} {\bibfnamefont {T.~A.}\
  \bibnamefont {Ely}},\ }\href@noop {} {\bibfield  {journal} {\bibinfo
  {journal} {Nature}\ }\textbf {\bibinfo {volume} {595}},\ \bibinfo {pages}
  {43} (\bibinfo {year} {2021})}\BibitemShut {NoStop}%
\bibitem [{\citenamefont {Brewer}\ \emph {et~al.}(2019)\citenamefont {Brewer},
  \citenamefont {Chen}, \citenamefont {Hankin}, \citenamefont {Clements},
  \citenamefont {Chou}, \citenamefont {Wineland}, \citenamefont {Hume},\ and\
  \citenamefont {Leibrandt}}]{Brewer2019-jh}%
  \BibitemOpen
  \bibfield  {author} {\bibinfo {author} {\bibfnamefont {S.~M.}\ \bibnamefont
  {Brewer}}, \bibinfo {author} {\bibfnamefont {J.-S.}\ \bibnamefont {Chen}},
  \bibinfo {author} {\bibfnamefont {A.~M.}\ \bibnamefont {Hankin}}, \bibinfo
  {author} {\bibfnamefont {E.~R.}\ \bibnamefont {Clements}}, \bibinfo {author}
  {\bibfnamefont {C.~W.}\ \bibnamefont {Chou}}, \bibinfo {author}
  {\bibfnamefont {D.~J.}\ \bibnamefont {Wineland}}, \bibinfo {author}
  {\bibfnamefont {D.~B.}\ \bibnamefont {Hume}},\ and\ \bibinfo {author}
  {\bibfnamefont {D.~R.}\ \bibnamefont {Leibrandt}},\ }\href@noop {} {\bibfield
   {journal} {\bibinfo  {journal} {Phys. Rev. Lett.}\ }\textbf {\bibinfo
  {volume} {123}},\ \bibinfo {pages} {033201} (\bibinfo {year}
  {2019})}\BibitemShut {NoStop}%
\bibitem [{\citenamefont {Connell}\ \emph {et~al.}(2017)\citenamefont
  {Connell}, \citenamefont {Ghadimi}, \citenamefont {Blums}, \citenamefont
  {Norton}, \citenamefont {Fisher}, \citenamefont {Amini}, \citenamefont
  {Volin}, \citenamefont {Hayden}, \citenamefont {Pai}, \citenamefont
  {Kielpinski}, \citenamefont {Lobino},\ and\ \citenamefont
  {Streed}}]{Connell2017-yg}%
  \BibitemOpen
  \bibfield  {author} {\bibinfo {author} {\bibfnamefont {S.~C.}\ \bibnamefont
  {Connell}}, \bibinfo {author} {\bibfnamefont {M.}~\bibnamefont {Ghadimi}},
  \bibinfo {author} {\bibfnamefont {V.}~\bibnamefont {Blums}}, \bibinfo
  {author} {\bibfnamefont {B.~G.}\ \bibnamefont {Norton}}, \bibinfo {author}
  {\bibfnamefont {P.}~\bibnamefont {Fisher}}, \bibinfo {author} {\bibfnamefont
  {J.~M.}\ \bibnamefont {Amini}}, \bibinfo {author} {\bibfnamefont
  {C.}~\bibnamefont {Volin}}, \bibinfo {author} {\bibfnamefont
  {H.}~\bibnamefont {Hayden}}, \bibinfo {author} {\bibfnamefont {C.~S.}\
  \bibnamefont {Pai}}, \bibinfo {author} {\bibfnamefont {D.}~\bibnamefont
  {Kielpinski}}, \bibinfo {author} {\bibfnamefont {M.}~\bibnamefont {Lobino}},\
  and\ \bibinfo {author} {\bibfnamefont {E.~W.}\ \bibnamefont {Streed}},\ }in\
  \href@noop {} {\emph {\bibinfo {booktitle} {Frontiers in Optics 2017}}}\
  (\bibinfo  {publisher} {Optical Society of America},\ \bibinfo {year}
  {2017})\ p.\ \bibinfo {pages} {FM3E.5}\BibitemShut {NoStop}%
\bibitem [{\citenamefont {{Matthias Keller, Birgit Lange, Kazuhiro Hayasaka,
  Wolfgang Lange \& Herbert
  Walther}}(2004)}]{Matthias_Keller_Birgit_Lange_Kazuhiro_Hayasaka_Wolfgang_Lange_Herbert_Walther2004-ib}%
  \BibitemOpen
  \bibfield  {author} {\bibinfo {author} {\bibnamefont {{Matthias Keller,
  Birgit Lange, Kazuhiro Hayasaka, Wolfgang Lange \& Herbert Walther}}},\
  }\href@noop {} {\bibfield  {journal} {\bibinfo  {journal} {Nature}\ }\textbf
  {\bibinfo {volume} {431}},\ \bibinfo {pages} {1075} (\bibinfo {year}
  {2004})}\BibitemShut {NoStop}%
\bibitem [{\citenamefont {Schupp}\ \emph {et~al.}(2021)\citenamefont {Schupp},
  \citenamefont {Krcmarsky}, \citenamefont {Krutyanskiy}, \citenamefont
  {Meraner}, \citenamefont {Northup},\ and\ \citenamefont
  {Lanyon}}]{Schupp2021-os}%
  \BibitemOpen
  \bibfield  {author} {\bibinfo {author} {\bibfnamefont {J.}~\bibnamefont
  {Schupp}}, \bibinfo {author} {\bibfnamefont {V.}~\bibnamefont {Krcmarsky}},
  \bibinfo {author} {\bibfnamefont {V.}~\bibnamefont {Krutyanskiy}}, \bibinfo
  {author} {\bibfnamefont {M.}~\bibnamefont {Meraner}}, \bibinfo {author}
  {\bibfnamefont {T.~E.}\ \bibnamefont {Northup}},\ and\ \bibinfo {author}
  {\bibfnamefont {B.~P.}\ \bibnamefont {Lanyon}},\ }\href@noop {} {\bibfield
  {journal} {\bibinfo  {journal} {PRX Quantum}\ }\textbf {\bibinfo {volume}
  {2}},\ \bibinfo {pages} {020331} (\bibinfo {year} {2021})}\BibitemShut
  {NoStop}%
\bibitem [{\citenamefont {Cirac}\ and\ \citenamefont
  {Zoller}(1995)}]{Cirac1995-bz}%
  \BibitemOpen
  \bibfield  {author} {\bibinfo {author} {\bibfnamefont {J.~I.}\ \bibnamefont
  {Cirac}}\ and\ \bibinfo {author} {\bibfnamefont {P.}~\bibnamefont {Zoller}},\
  }\href@noop {} {\bibfield  {journal} {\bibinfo  {journal} {Phys. Rev. Lett.}\
  }\textbf {\bibinfo {volume} {74}},\ \bibinfo {pages} {4091} (\bibinfo {year}
  {1995})}\BibitemShut {NoStop}%
\bibitem [{\citenamefont {Brown}\ \emph {et~al.}(2016)\citenamefont {Brown},
  \citenamefont {Kim},\ and\ \citenamefont {Monroe}}]{Brown2016-me}%
  \BibitemOpen
  \bibfield  {author} {\bibinfo {author} {\bibfnamefont {K.~R.}\ \bibnamefont
  {Brown}}, \bibinfo {author} {\bibfnamefont {J.}~\bibnamefont {Kim}},\ and\
  \bibinfo {author} {\bibfnamefont {C.}~\bibnamefont {Monroe}},\ }\href@noop {}
  {\bibfield  {journal} {\bibinfo  {journal} {npj Quantum Information}\
  }\textbf {\bibinfo {volume} {2}},\ \bibinfo {pages} {1} (\bibinfo {year}
  {2016})}\BibitemShut {NoStop}%
\bibitem [{\citenamefont {M{\o}lmer}\ and\ \citenamefont
  {S{\o}rensen}(1999)}]{Molmer1999-dq}%
  \BibitemOpen
  \bibfield  {author} {\bibinfo {author} {\bibfnamefont {K.}~\bibnamefont
  {M{\o}lmer}}\ and\ \bibinfo {author} {\bibfnamefont {A.}~\bibnamefont
  {S{\o}rensen}},\ }\href@noop {} {\bibfield  {journal} {\bibinfo  {journal}
  {Phys. Rev. Lett.}\ }\textbf {\bibinfo {volume} {82}},\ \bibinfo {pages}
  {1835} (\bibinfo {year} {1999})}\BibitemShut {NoStop}%
\bibitem [{\citenamefont {Gaebler}\ \emph {et~al.}(2016)\citenamefont
  {Gaebler}, \citenamefont {Tan}, \citenamefont {Lin}, \citenamefont {Wan},
  \citenamefont {Bowler}, \citenamefont {Keith}, \citenamefont {Glancy},
  \citenamefont {Coakley}, \citenamefont {Knill}, \citenamefont {Leibfried},\
  and\ \citenamefont {Wineland}}]{Gaebler2016-rd}%
  \BibitemOpen
  \bibfield  {author} {\bibinfo {author} {\bibfnamefont {J.~P.}\ \bibnamefont
  {Gaebler}}, \bibinfo {author} {\bibfnamefont {T.~R.}\ \bibnamefont {Tan}},
  \bibinfo {author} {\bibfnamefont {Y.}~\bibnamefont {Lin}}, \bibinfo {author}
  {\bibfnamefont {Y.}~\bibnamefont {Wan}}, \bibinfo {author} {\bibfnamefont
  {R.}~\bibnamefont {Bowler}}, \bibinfo {author} {\bibfnamefont {A.~C.}\
  \bibnamefont {Keith}}, \bibinfo {author} {\bibfnamefont {S.}~\bibnamefont
  {Glancy}}, \bibinfo {author} {\bibfnamefont {K.}~\bibnamefont {Coakley}},
  \bibinfo {author} {\bibfnamefont {E.}~\bibnamefont {Knill}}, \bibinfo
  {author} {\bibfnamefont {D.}~\bibnamefont {Leibfried}},\ and\ \bibinfo
  {author} {\bibfnamefont {D.~J.}\ \bibnamefont {Wineland}},\ }\href@noop {}
  {\bibfield  {journal} {\bibinfo  {journal} {Phys. Rev. Lett.}\ }\textbf
  {\bibinfo {volume} {117}},\ \bibinfo {pages} {060505} (\bibinfo {year}
  {2016})}\BibitemShut {NoStop}%
\bibitem [{\citenamefont {Ospelkaus}\ \emph {et~al.}(2011)\citenamefont
  {Ospelkaus}, \citenamefont {Warring}, \citenamefont {Colombe}, \citenamefont
  {Brown}, \citenamefont {Amini}, \citenamefont {Leibfried},\ and\
  \citenamefont {Wineland}}]{Ospelkaus2011-go}%
  \BibitemOpen
  \bibfield  {author} {\bibinfo {author} {\bibfnamefont {C.}~\bibnamefont
  {Ospelkaus}}, \bibinfo {author} {\bibfnamefont {U.}~\bibnamefont {Warring}},
  \bibinfo {author} {\bibfnamefont {Y.}~\bibnamefont {Colombe}}, \bibinfo
  {author} {\bibfnamefont {K.~R.}\ \bibnamefont {Brown}}, \bibinfo {author}
  {\bibfnamefont {J.~M.}\ \bibnamefont {Amini}}, \bibinfo {author}
  {\bibfnamefont {D.}~\bibnamefont {Leibfried}},\ and\ \bibinfo {author}
  {\bibfnamefont {D.~J.}\ \bibnamefont {Wineland}},\ }\href@noop {} {\bibfield
  {journal} {\bibinfo  {journal} {Nature}\ }\textbf {\bibinfo {volume} {476}},\
  \bibinfo {pages} {181} (\bibinfo {year} {2011})}\BibitemShut {NoStop}%
\bibitem [{\citenamefont {Gottesman}\ \emph {et~al.}(2001)\citenamefont
  {Gottesman}, \citenamefont {Kitaev},\ and\ \citenamefont
  {Preskill}}]{Gottesman2001-rn}%
  \BibitemOpen
  \bibfield  {author} {\bibinfo {author} {\bibfnamefont {D.}~\bibnamefont
  {Gottesman}}, \bibinfo {author} {\bibfnamefont {A.}~\bibnamefont {Kitaev}},\
  and\ \bibinfo {author} {\bibfnamefont {J.}~\bibnamefont {Preskill}},\
  }\href@noop {} {\bibfield  {journal} {\bibinfo  {journal} {Phys. Rev. A}\
  }\textbf {\bibinfo {volume} {64}},\ \bibinfo {pages} {012310} (\bibinfo
  {year} {2001})}\BibitemShut {NoStop}%
\bibitem [{\citenamefont {Fl{\"u}hmann}\ \emph {et~al.}(2019)\citenamefont
  {Fl{\"u}hmann}, \citenamefont {Nguyen}, \citenamefont {Marinelli},
  \citenamefont {Negnevitsky}, \citenamefont {Mehta},\ and\ \citenamefont
  {Home}}]{Fluhmann2019-ki}%
  \BibitemOpen
  \bibfield  {author} {\bibinfo {author} {\bibfnamefont {C.}~\bibnamefont
  {Fl{\"u}hmann}}, \bibinfo {author} {\bibfnamefont {T.~L.}\ \bibnamefont
  {Nguyen}}, \bibinfo {author} {\bibfnamefont {M.}~\bibnamefont {Marinelli}},
  \bibinfo {author} {\bibfnamefont {V.}~\bibnamefont {Negnevitsky}}, \bibinfo
  {author} {\bibfnamefont {K.}~\bibnamefont {Mehta}},\ and\ \bibinfo {author}
  {\bibfnamefont {J.~P.}\ \bibnamefont {Home}},\ }\href@noop {} {\bibfield
  {journal} {\bibinfo  {journal} {Nature}\ }\textbf {\bibinfo {volume} {566}},\
  \bibinfo {pages} {513} (\bibinfo {year} {2019})}\BibitemShut {NoStop}%
\bibitem [{\citenamefont {Brownnutt}\ \emph {et~al.}(2015)\citenamefont
  {Brownnutt}, \citenamefont {Kumph}, \citenamefont {Rabl},\ and\ \citenamefont
  {Blatt}}]{Brownnutt2015-oq}%
  \BibitemOpen
  \bibfield  {author} {\bibinfo {author} {\bibfnamefont {M.}~\bibnamefont
  {Brownnutt}}, \bibinfo {author} {\bibfnamefont {M.}~\bibnamefont {Kumph}},
  \bibinfo {author} {\bibfnamefont {P.}~\bibnamefont {Rabl}},\ and\ \bibinfo
  {author} {\bibfnamefont {R.}~\bibnamefont {Blatt}},\ }\href@noop {}
  {\bibfield  {journal} {\bibinfo  {journal} {Rev. Mod. Phys.}\ }\textbf
  {\bibinfo {volume} {87}},\ \bibinfo {pages} {1419} (\bibinfo {year}
  {2015})}\BibitemShut {NoStop}%
\bibitem [{\citenamefont {Schioppo}\ \emph {et~al.}(2012)\citenamefont
  {Schioppo}, \citenamefont {Poli}, \citenamefont {Prevedelli}, \citenamefont
  {Falke}, \citenamefont {Lisdat}, \citenamefont {Sterr},\ and\ \citenamefont
  {Tino}}]{Schioppo2012-cf}%
  \BibitemOpen
  \bibfield  {author} {\bibinfo {author} {\bibfnamefont {M.}~\bibnamefont
  {Schioppo}}, \bibinfo {author} {\bibfnamefont {N.}~\bibnamefont {Poli}},
  \bibinfo {author} {\bibfnamefont {M.}~\bibnamefont {Prevedelli}}, \bibinfo
  {author} {\bibfnamefont {S.}~\bibnamefont {Falke}}, \bibinfo {author}
  {\bibfnamefont {C.}~\bibnamefont {Lisdat}}, \bibinfo {author} {\bibfnamefont
  {U.}~\bibnamefont {Sterr}},\ and\ \bibinfo {author} {\bibfnamefont {G.~M.}\
  \bibnamefont {Tino}},\ }\href@noop {} {\bibfield  {journal} {\bibinfo
  {journal} {Rev. Sci. Instrum.}\ }\textbf {\bibinfo {volume} {83}},\ \bibinfo
  {pages} {103101} (\bibinfo {year} {2012})}\BibitemShut {NoStop}%
\bibitem [{\citenamefont {Olmschenk}\ and\ \citenamefont
  {Becker}(2017)}]{Olmschenk2017-mf}%
  \BibitemOpen
  \bibfield  {author} {\bibinfo {author} {\bibfnamefont {S.}~\bibnamefont
  {Olmschenk}}\ and\ \bibinfo {author} {\bibfnamefont {P.}~\bibnamefont
  {Becker}},\ }\href@noop {} {\bibfield  {journal} {\bibinfo  {journal} {Appl.
  Phys. B}\ }\textbf {\bibinfo {volume} {123}},\ \bibinfo {pages} {99}
  (\bibinfo {year} {2017})}\BibitemShut {NoStop}%
\bibitem [{\citenamefont {Piotrowski}\ \emph {et~al.}(2020)\citenamefont
  {Piotrowski}, \citenamefont {Scarabel}, \citenamefont {Lobino}, \citenamefont
  {Streed},\ and\ \citenamefont {Gensemer}}]{Piotrowski2020-kz}%
  \BibitemOpen
  \bibfield  {author} {\bibinfo {author} {\bibfnamefont {M.}~\bibnamefont
  {Piotrowski}}, \bibinfo {author} {\bibfnamefont {J.}~\bibnamefont
  {Scarabel}}, \bibinfo {author} {\bibfnamefont {M.}~\bibnamefont {Lobino}},
  \bibinfo {author} {\bibfnamefont {E.}~\bibnamefont {Streed}},\ and\ \bibinfo
  {author} {\bibfnamefont {S.}~\bibnamefont {Gensemer}},\ }\href@noop {}
  {\bibfield  {journal} {\bibinfo  {journal} {OSA Continuum, OSAC}\ }\textbf
  {\bibinfo {volume} {3}},\ \bibinfo {pages} {2210} (\bibinfo {year}
  {2020})}\BibitemShut {NoStop}%
\bibitem [{\citenamefont {Zimmermann}\ \emph {et~al.}(2012)\citenamefont
  {Zimmermann}, \citenamefont {Okhapkin}, \citenamefont {Herrera-Sancho},\ and\
  \citenamefont {Peik}}]{Zimmermann2012-gv}%
  \BibitemOpen
  \bibfield  {author} {\bibinfo {author} {\bibfnamefont {K.}~\bibnamefont
  {Zimmermann}}, \bibinfo {author} {\bibfnamefont {M.~V.}\ \bibnamefont
  {Okhapkin}}, \bibinfo {author} {\bibfnamefont {O.~A.}\ \bibnamefont
  {Herrera-Sancho}},\ and\ \bibinfo {author} {\bibfnamefont {E.}~\bibnamefont
  {Peik}},\ }\href@noop {} {\bibfield  {journal} {\bibinfo  {journal} {Appl.
  Phys. B}\ }\textbf {\bibinfo {volume} {107}},\ \bibinfo {pages} {883}
  (\bibinfo {year} {2012})}\BibitemShut {NoStop}%
\bibitem [{\citenamefont {Sheridan}\ \emph {et~al.}(2011)\citenamefont
  {Sheridan}, \citenamefont {Lange},\ and\ \citenamefont
  {Keller}}]{Sheridan2011-ck}%
  \BibitemOpen
  \bibfield  {author} {\bibinfo {author} {\bibfnamefont {K.}~\bibnamefont
  {Sheridan}}, \bibinfo {author} {\bibfnamefont {W.}~\bibnamefont {Lange}},\
  and\ \bibinfo {author} {\bibfnamefont {M.}~\bibnamefont {Keller}},\
  }\href@noop {} {\bibfield  {journal} {\bibinfo  {journal} {Appl. Phys. B}\
  }\textbf {\bibinfo {volume} {104}},\ \bibinfo {pages} {755} (\bibinfo {year}
  {2011})}\BibitemShut {NoStop}%
\bibitem [{\citenamefont {Shao}\ \emph {et~al.}(2018)\citenamefont {Shao},
  \citenamefont {Wang}, \citenamefont {Zeng}, \citenamefont {Guan},\ and\
  \citenamefont {Gao}}]{Shao2018-ak}%
  \BibitemOpen
  \bibfield  {author} {\bibinfo {author} {\bibfnamefont {H.}~\bibnamefont
  {Shao}}, \bibinfo {author} {\bibfnamefont {M.}~\bibnamefont {Wang}}, \bibinfo
  {author} {\bibfnamefont {M.}~\bibnamefont {Zeng}}, \bibinfo {author}
  {\bibfnamefont {H.}~\bibnamefont {Guan}},\ and\ \bibinfo {author}
  {\bibfnamefont {K.}~\bibnamefont {Gao}},\ }\href@noop {} {\bibfield
  {journal} {\bibinfo  {journal} {J. Phys. Commun.}\ }\textbf {\bibinfo
  {volume} {2}},\ \bibinfo {pages} {095019} (\bibinfo {year}
  {2018})}\BibitemShut {NoStop}%
\bibitem [{\citenamefont {Sameed}\ \emph {et~al.}(2020)\citenamefont {Sameed},
  \citenamefont {Maxwell},\ and\ \citenamefont {Madsen}}]{Sameed2020-ac}%
  \BibitemOpen
  \bibfield  {author} {\bibinfo {author} {\bibfnamefont {M.}~\bibnamefont
  {Sameed}}, \bibinfo {author} {\bibfnamefont {D.}~\bibnamefont {Maxwell}},\
  and\ \bibinfo {author} {\bibfnamefont {N.}~\bibnamefont {Madsen}},\
  }\href@noop {} {\bibfield  {journal} {\bibinfo  {journal} {New J. Phys.}\
  }\textbf {\bibinfo {volume} {22}},\ \bibinfo {pages} {013009} (\bibinfo
  {year} {2020})}\BibitemShut {NoStop}%
\bibitem [{\citenamefont {Leibrandt}\ \emph {et~al.}(2007)\citenamefont
  {Leibrandt}, \citenamefont {Clark}, \citenamefont {Labaziewicz},
  \citenamefont {Antohi}, \citenamefont {Bakr}, \citenamefont {Brown},\ and\
  \citenamefont {Chuang}}]{Leibrandt2007-au}%
  \BibitemOpen
  \bibfield  {author} {\bibinfo {author} {\bibfnamefont {D.~R.}\ \bibnamefont
  {Leibrandt}}, \bibinfo {author} {\bibfnamefont {R.~J.}\ \bibnamefont
  {Clark}}, \bibinfo {author} {\bibfnamefont {J.}~\bibnamefont {Labaziewicz}},
  \bibinfo {author} {\bibfnamefont {P.}~\bibnamefont {Antohi}}, \bibinfo
  {author} {\bibfnamefont {W.}~\bibnamefont {Bakr}}, \bibinfo {author}
  {\bibfnamefont {K.~R.}\ \bibnamefont {Brown}},\ and\ \bibinfo {author}
  {\bibfnamefont {I.~L.}\ \bibnamefont {Chuang}},\ }\href@noop {} {\bibfield
  {journal} {\bibinfo  {journal} {Phys. Rev. A}\ }\textbf {\bibinfo {volume}
  {76}},\ \bibinfo {pages} {055403} (\bibinfo {year} {2007})}\BibitemShut
  {NoStop}%
\bibitem [{\citenamefont {Vrijsen}\ \emph {et~al.}(2019)\citenamefont
  {Vrijsen}, \citenamefont {Aikyo}, \citenamefont {Spivey}, \citenamefont
  {Inlek},\ and\ \citenamefont {Kim}}]{Vrijsen2019-xr}%
  \BibitemOpen
  \bibfield  {author} {\bibinfo {author} {\bibfnamefont {G.}~\bibnamefont
  {Vrijsen}}, \bibinfo {author} {\bibfnamefont {Y.}~\bibnamefont {Aikyo}},
  \bibinfo {author} {\bibfnamefont {R.~F.}\ \bibnamefont {Spivey}}, \bibinfo
  {author} {\bibfnamefont {I.~V.}\ \bibnamefont {Inlek}},\ and\ \bibinfo
  {author} {\bibfnamefont {J.}~\bibnamefont {Kim}},\ }\href@noop {} {\bibfield
  {journal} {\bibinfo  {journal} {Opt. Express}\ }\textbf {\bibinfo {volume}
  {27}},\ \bibinfo {pages} {33907} (\bibinfo {year} {2019})}\BibitemShut
  {NoStop}%
\bibitem [{\citenamefont {Wu}\ \emph {et~al.}(2021)\citenamefont {Wu},
  \citenamefont {Filzinger}, \citenamefont {Shi}, \citenamefont {Wang},\ and\
  \citenamefont {Zhang}}]{Wu2021-nq}%
  \BibitemOpen
  \bibfield  {author} {\bibinfo {author} {\bibfnamefont {Q.}~\bibnamefont
  {Wu}}, \bibinfo {author} {\bibfnamefont {M.}~\bibnamefont {Filzinger}},
  \bibinfo {author} {\bibfnamefont {Y.}~\bibnamefont {Shi}}, \bibinfo {author}
  {\bibfnamefont {Z.}~\bibnamefont {Wang}},\ and\ \bibinfo {author}
  {\bibfnamefont {J.}~\bibnamefont {Zhang}},\ }\href@noop {} {\bibfield
  {journal} {\bibinfo  {journal} {Rev. Sci. Instrum.}\ }\textbf {\bibinfo
  {volume} {92}},\ \bibinfo {pages} {063201} (\bibinfo {year}
  {2021})}\BibitemShut {NoStop}%
\bibitem [{\citenamefont {Kielpinski}\ \emph {et~al.}(2002)\citenamefont
  {Kielpinski}, \citenamefont {Monroe},\ and\ \citenamefont
  {Wineland}}]{Kielpinski2002-dy}%
  \BibitemOpen
  \bibfield  {author} {\bibinfo {author} {\bibfnamefont {D.}~\bibnamefont
  {Kielpinski}}, \bibinfo {author} {\bibfnamefont {C.}~\bibnamefont {Monroe}},\
  and\ \bibinfo {author} {\bibfnamefont {D.~J.}\ \bibnamefont {Wineland}},\
  }\href@noop {} {\bibfield  {journal} {\bibinfo  {journal} {Nature}\ }\textbf
  {\bibinfo {volume} {417}},\ \bibinfo {pages} {709} (\bibinfo {year}
  {2002})}\BibitemShut {NoStop}%
\bibitem [{\citenamefont {Hill}\ and\ \citenamefont
  {McClelland}(2004)}]{Hill2004-vv}%
  \BibitemOpen
  \bibfield  {author} {\bibinfo {author} {\bibfnamefont {S.~B.}\ \bibnamefont
  {Hill}}\ and\ \bibinfo {author} {\bibfnamefont {J.~J.}\ \bibnamefont
  {McClelland}},\ }\href@noop {} {\bibfield  {journal} {\bibinfo  {journal} {J.
  Opt. Soc. Am. B, JOSAB}\ }\textbf {\bibinfo {volume} {21}},\ \bibinfo {pages}
  {473} (\bibinfo {year} {2004})}\BibitemShut {NoStop}%
\bibitem [{\citenamefont {Schnitzler}\ \emph {et~al.}(2009)\citenamefont
  {Schnitzler}, \citenamefont {Linke}, \citenamefont {Fickler}, \citenamefont
  {Meijer}, \citenamefont {Schmidt-Kaler},\ and\ \citenamefont
  {Singer}}]{Schnitzler2009-qa}%
  \BibitemOpen
  \bibfield  {author} {\bibinfo {author} {\bibfnamefont {W.}~\bibnamefont
  {Schnitzler}}, \bibinfo {author} {\bibfnamefont {N.~M.}\ \bibnamefont
  {Linke}}, \bibinfo {author} {\bibfnamefont {R.}~\bibnamefont {Fickler}},
  \bibinfo {author} {\bibfnamefont {J.}~\bibnamefont {Meijer}}, \bibinfo
  {author} {\bibfnamefont {F.}~\bibnamefont {Schmidt-Kaler}},\ and\ \bibinfo
  {author} {\bibfnamefont {K.}~\bibnamefont {Singer}},\ }\href@noop {}
  {\bibfield  {journal} {\bibinfo  {journal} {Phys. Rev. Lett.}\ }\textbf
  {\bibinfo {volume} {102}},\ \bibinfo {pages} {070501} (\bibinfo {year}
  {2009})}\BibitemShut {NoStop}%
\bibitem [{\citenamefont {Schnitzler}\ \emph {et~al.}(2010)\citenamefont
  {Schnitzler}, \citenamefont {Jacob}, \citenamefont {Fickler}, \citenamefont
  {Schmidt-Kaler},\ and\ \citenamefont {Singer}}]{Schnitzler2010-op}%
  \BibitemOpen
  \bibfield  {author} {\bibinfo {author} {\bibfnamefont {W.}~\bibnamefont
  {Schnitzler}}, \bibinfo {author} {\bibfnamefont {G.}~\bibnamefont {Jacob}},
  \bibinfo {author} {\bibfnamefont {R.}~\bibnamefont {Fickler}}, \bibinfo
  {author} {\bibfnamefont {F.}~\bibnamefont {Schmidt-Kaler}},\ and\ \bibinfo
  {author} {\bibfnamefont {K.}~\bibnamefont {Singer}},\ }\href@noop {}
  {\bibfield  {journal} {\bibinfo  {journal} {New J. Phys.}\ }\textbf {\bibinfo
  {volume} {12}},\ \bibinfo {pages} {065023} (\bibinfo {year}
  {2010})}\BibitemShut {NoStop}%
\bibitem [{\citenamefont {Hanssen}\ \emph {et~al.}(2006)\citenamefont
  {Hanssen}, \citenamefont {McClelland}, \citenamefont {Dakin},\ and\
  \citenamefont {Jacka}}]{Hanssen2006-te}%
  \BibitemOpen
  \bibfield  {author} {\bibinfo {author} {\bibfnamefont {J.~L.}\ \bibnamefont
  {Hanssen}}, \bibinfo {author} {\bibfnamefont {J.~J.}\ \bibnamefont
  {McClelland}}, \bibinfo {author} {\bibfnamefont {E.~A.}\ \bibnamefont
  {Dakin}},\ and\ \bibinfo {author} {\bibfnamefont {M.}~\bibnamefont {Jacka}},\
  }\href@noop {} {\bibfield  {journal} {\bibinfo  {journal} {Phys. Rev. A}\
  }\textbf {\bibinfo {volume} {74}},\ \bibinfo {pages} {063416} (\bibinfo
  {year} {2006})}\BibitemShut {NoStop}%
\bibitem [{\citenamefont {Niemann}\ \emph {et~al.}(2019)\citenamefont
  {Niemann}, \citenamefont {Meiners}, \citenamefont {Mielke}, \citenamefont
  {Borchert}, \citenamefont {Cornejo}, \citenamefont {Ulmer},\ and\
  \citenamefont {Ospelkaus}}]{Niemann2019-yk}%
  \BibitemOpen
  \bibfield  {author} {\bibinfo {author} {\bibfnamefont {M.}~\bibnamefont
  {Niemann}}, \bibinfo {author} {\bibfnamefont {T.}~\bibnamefont {Meiners}},
  \bibinfo {author} {\bibfnamefont {J.}~\bibnamefont {Mielke}}, \bibinfo
  {author} {\bibfnamefont {M.~J.}\ \bibnamefont {Borchert}}, \bibinfo {author}
  {\bibfnamefont {J.~M.}\ \bibnamefont {Cornejo}}, \bibinfo {author}
  {\bibfnamefont {S.}~\bibnamefont {Ulmer}},\ and\ \bibinfo {author}
  {\bibfnamefont {C.}~\bibnamefont {Ospelkaus}},\ }\href@noop {} {\bibfield
  {journal} {\bibinfo  {journal} {Meas. Sci. Technol.}\ }\textbf {\bibinfo
  {volume} {31}},\ \bibinfo {pages} {035003} (\bibinfo {year}
  {2019})}\BibitemShut {NoStop}%
\bibitem [{\citenamefont {Dubielzig}\ \emph {et~al.}(2021)\citenamefont
  {Dubielzig}, \citenamefont {Halama}, \citenamefont {Hahn}, \citenamefont
  {Zarantonello}, \citenamefont {Niemann}, \citenamefont {Bautista-Salvador},\
  and\ \citenamefont {Ospelkaus}}]{Dubielzig2021-qh}%
  \BibitemOpen
  \bibfield  {author} {\bibinfo {author} {\bibfnamefont {T.}~\bibnamefont
  {Dubielzig}}, \bibinfo {author} {\bibfnamefont {S.}~\bibnamefont {Halama}},
  \bibinfo {author} {\bibfnamefont {H.}~\bibnamefont {Hahn}}, \bibinfo {author}
  {\bibfnamefont {G.}~\bibnamefont {Zarantonello}}, \bibinfo {author}
  {\bibfnamefont {M.}~\bibnamefont {Niemann}}, \bibinfo {author} {\bibfnamefont
  {A.}~\bibnamefont {Bautista-Salvador}},\ and\ \bibinfo {author}
  {\bibfnamefont {C.}~\bibnamefont {Ospelkaus}},\ }\href@noop {} {\bibfield
  {journal} {\bibinfo  {journal} {Rev. Sci. Instrum.}\ }\textbf {\bibinfo
  {volume} {92}},\ \bibinfo {pages} {043201} (\bibinfo {year}
  {2021})}\BibitemShut {NoStop}%
\bibitem [{\citenamefont {Hong}\ \emph {et~al.}(2016)\citenamefont {Hong},
  \citenamefont {Lee}, \citenamefont {Cheon}, \citenamefont {Kim},\ and\
  \citenamefont {Cho}}]{Hong2016-mp}%
  \BibitemOpen
  \bibfield  {author} {\bibinfo {author} {\bibfnamefont {S.}~\bibnamefont
  {Hong}}, \bibinfo {author} {\bibfnamefont {M.}~\bibnamefont {Lee}}, \bibinfo
  {author} {\bibfnamefont {H.}~\bibnamefont {Cheon}}, \bibinfo {author}
  {\bibfnamefont {T.}~\bibnamefont {Kim}},\ and\ \bibinfo {author}
  {\bibfnamefont {D.-I.~D.}\ \bibnamefont {Cho}},\ }\href@noop {} {\bibfield
  {journal} {\bibinfo  {journal} {Sensors}\ }\textbf {\bibinfo {volume} {16}},\
  \bibinfo {pages} {616} (\bibinfo {year} {2016})}\BibitemShut {NoStop}%
\bibitem [{\citenamefont {Jones}\ \emph {et~al.}(1989)\citenamefont {Jones},
  \citenamefont {Braunlich}, \citenamefont {Thomas~Casper}, \citenamefont
  {Shen},\ and\ \citenamefont {Kelly}}]{Jones1989-hw}%
  \BibitemOpen
  \bibfield  {author} {\bibinfo {author} {\bibfnamefont {S.~C.}\ \bibnamefont
  {Jones}}, \bibinfo {author} {\bibfnamefont {P.}~\bibnamefont {Braunlich}},
  \bibinfo {author} {\bibfnamefont {R.}~\bibnamefont {Thomas~Casper}}, \bibinfo
  {author} {\bibfnamefont {X.-A.}\ \bibnamefont {Shen}},\ and\ \bibinfo
  {author} {\bibfnamefont {P.}~\bibnamefont {Kelly}},\ }\href@noop {}
  {\bibfield  {journal} {\bibinfo  {journal} {Optical Engineering}\ }\textbf
  {\bibinfo {volume} {28}},\ \bibinfo {pages} {1039} (\bibinfo {year}
  {1989})}\BibitemShut {NoStop}%
\end{thebibliography}%
\end{document}